\documentclass[ 
    ,final            
  ] 
  {aipproc} 
 
\layoutstyle{6x9} 
 
\begin{document} 
 
\title{1RXS\,J180834.7+101041 is a new cataclysmic variable  
with non-uniform disc } 
 
\classification{97.10.Gz,97.80.Gm,97.80.Fk,97.80.Hn} 
\keywords      { binaries: eclipsing, binaries: spectroscopic,  novae, cataclysmic variables,  
 stars: individual: 1RXS\,J180834.7+101041} 
 
\author{D. G. Yakin}{ 
  address={Kazan Federal University, Kremlevskaya str. 18, 42008 Kazan, Russia} 
} 
 
\author{V. F. Suleimanov}{ 
  address={Insitute for Astronomy and Astrophysics, Kepler Center for Astro and Particle Physics, Eberhard Karls University, Sand 1, 72076 T\"ubingen, Germany}, 
altaddress={Kazan Federal University, Kremlevskaya str. 18, 42008 Kazan, Russia}, 
email={suleimanov@astro.uni-tuebingen.de} 
} 
 
\author{V. V. Shimansky}{ 
  address={Kazan Federal University, Kremlevskaya str. 18, 42008 Kazan, Russia} 
} 
\author{N. V. Borisov}{ 
  address={Special Astrophysical Observatory, 369167 N. Arkhyz, Russia} 
} 
\author{I.~F.~Bikmaev }{ 
  address={Kazan Federal University, Kremlevskaya str. 18, 42008 Kazan, Russia} 
,altaddress={ Tatarstan Academy of Sciences, Baumana str. 20, 420111 Kazan, Russia}, 
} 
\author{N. A. Sakhibullin}{ 
  address={Kazan Federal University, Kremlevskaya str. 18, 42008 Kazan, Russia} 
,altaddress={ Tatarstan Academy of Sciences, Baumana str. 20, 420111 Kazan, Russia}, 
}

\begin{abstract} 
  
Results of photometric and spectroscopic investigations of the recently 
discovered disc cataclysmic variable star 1RXS\,J180834.7+101041 are 
presented. 
Emission spectra of the system show broad double peaked hydrogen and 
helium emission lines. Doppler maps for the hydrogen lines demonstrate strongly 
non-uniform emissivity distribution in the disc, similar to that found in IP 
Peg. It means that the system is a new cataclysmic variable with a spiral 
density wave in the disc. Masses of the components ($M_{\rm WD} = 0.8 \pm 0.22 M_{\odot}$ and 
$M_{\rm RD} = 0.14 \pm 0.02~ M_{\odot}$), and the orbit inclination   
($i = 78^{\circ} \pm 1.^{\circ}5$) were estimated  using the various  
well-known relations for cataclysmic variables. 
 
\end{abstract} 
 
\maketitle 
 
 
\section{Introduction} 
 
1RXS\,J180834.7+101041 = USNO-B1 1001-0317189 
($\alpha_{2000}=18^{h}08^{m}35^{s}.8$, 
$\delta_{2000}=+10^{\circ}10'30''.2$), later 1RXS\,J1808, is a {\it ROSAT} 
X-ray source, recently identified as an eclipsing cataclysmic variable (CV) \citep{Den:08,Bikm:08}. 
Using the observed eclipses, \cite{Den:08} found the orbital period  
of this close binary system  (0.$^d$070037(1)) from the photometric observations. 
They also found that the brightness of the object varies 
with an amplitude $\sim 1^m$ on a few weeks time scale and classified this CV as a polar. 
But a spectrum of this system \citep{Bikm:08} shows the double peaked hydrogen and helium 
emission lines. Therefore, 1RXS\,J1808 contains an accretion disc around the white dwarf 
and, therefore, can not be a polar. 
 
Here we present new spectroscopic observations of this CV and their preliminary 
analysis together with the analysis of avaliable photometric data.

\section{Observations} 
 
\begin{figure} 
  \includegraphics[height=.23\textheight]{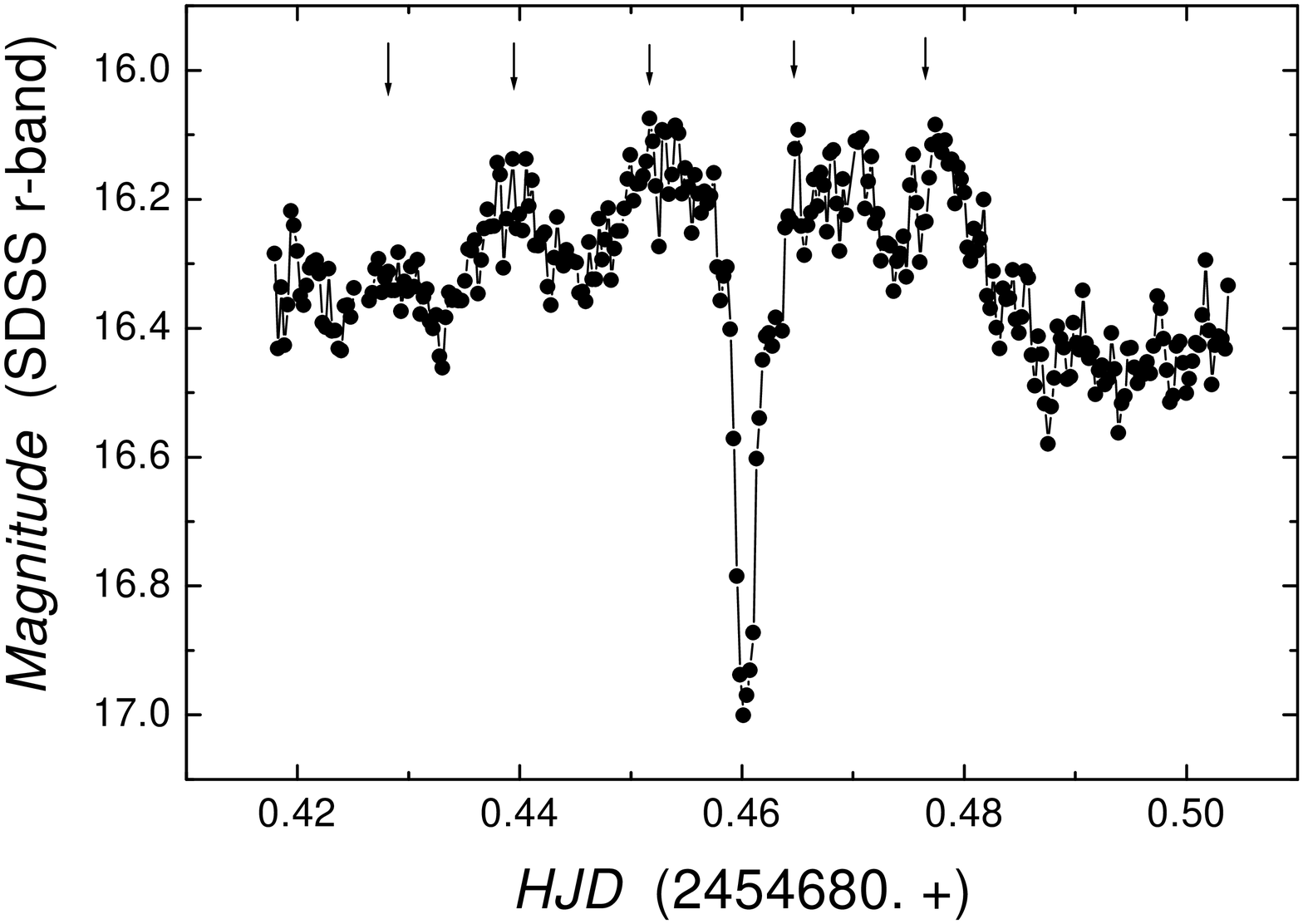} 
  \includegraphics[height=.23\textheight]{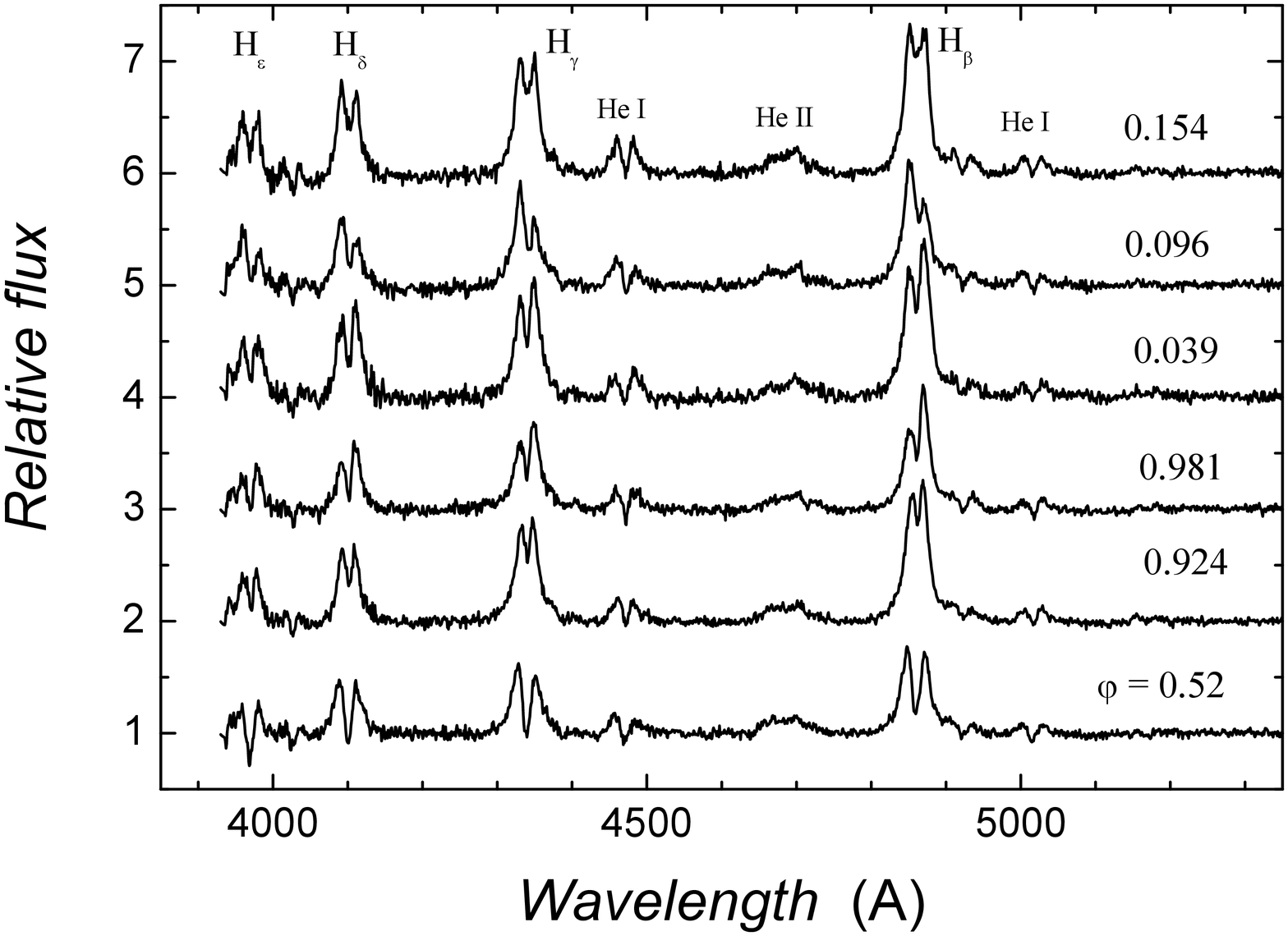} 
  \caption{ \label{v3sfig1} 
{\it Left:} Observed light curve in SDSS r-band. Maxima of the quasiperiodic brightness 
variations with the period 
$P \approx 0.2~ P_{\rm orb}$ are marked by arrows. {\it Right:} Normalized observed spectra in six orbital phases.} 
\end{figure}

The photometric observations of 1RXS\,J1808 were performed on 
August 1-2 2008, with the  1.5--meter Russian-Turkish 
telescope RTT-150 at the TUBITAK National Observatory (Turkey).  
Observations were 
carried out in the SDSS r-band with an exposure time of 25 sec. 
The total observation 
time  was about 2 hours and calibrated against SDSS standard stars. 
The obtained light curve is shown in Fig.\,\ref{v3sfig1}\,(left panel). 
Spectroscopic observations of 1RXS\,J1808 were carried out on August 9-10, 2008, by the 6--meter telescope BTA of the Special Astrophysical Observatory with 
the SCORPIO focal reducer \citep{AM:05}, which gives a  $\Delta\lambda$ = 5.0 
\AA~ resolution in the wavelength region  $\Delta \lambda$ 3900--5700 
\AA. 16 subsequent spectra  with the same exposure time of 300 s 
and signal-to noise ratio  $S/N \approx$ 55 - 65 were obtained.  
Examples of the obtained spectra are shown in Fig.\,\ref{v3sfig1}\,(right panel). 
The orbital phases are counted from the photometric minimum. 
 
\section{Data analysis} 
 
The most prominent details in the observed spectra of 1RXS\,J1808 are the broad double  
peaked emission lines of 
hydrogen as well as  neutral and ionized helium (Fig.~\ref{v3sfig1}\,right panel). The line profiles are strongly 
changing  with orbital phase, especially at the moment of the eclipse. 
 
The radial velocity of the white dwarf was measured using the hydrogen emission lines by Shafter's method \citep{sh:83}.  
Dependence of the radial velocity on the orbital phase for H$_{\beta}$ is 
shown in Fig.~\ref{v3sfig3}\,(left panel). Surprisingly the radial velocity maximum is at the phase 0 (at the eclipse) instead of 0.75. 
 This huge phase shift means that the radial velocity curve 
reflects the orbital motion of the bright spot rather than the motion of the white dwarf itself.  
By the way, the radial velocity curves were fitted by sine in the phase 
range 0.2 -- 0.9, and a value $K_1$ = 70 $\pm$ 10 km s$^{-1}$ was obtained. 
 
\begin{figure} 
\includegraphics[bb= 45 405 345 735,clip,angle=90,height=0.3\textheight]{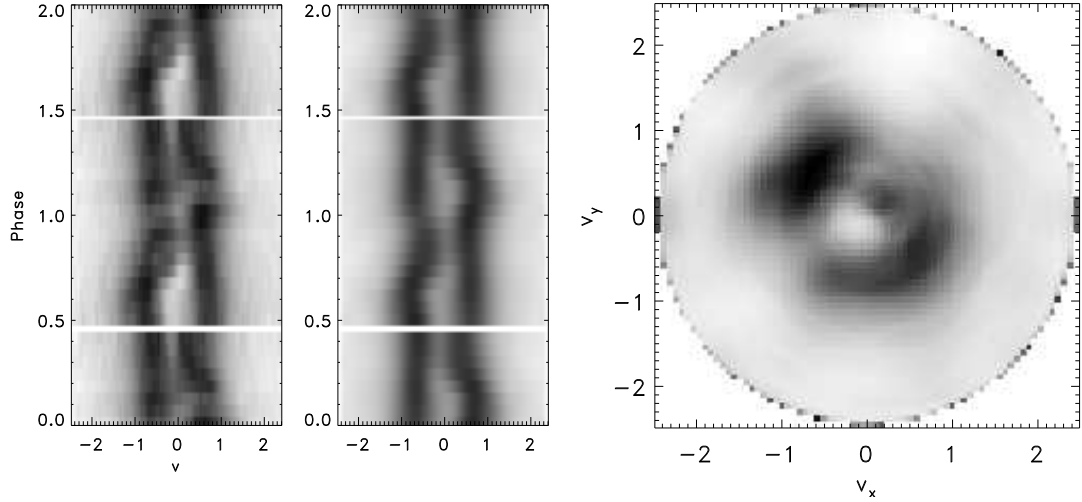} 
\includegraphics[bb= 347 495 565 738,clip,angle=90,height=0.3\textheight]{v3suleimanov02.eps} 
  \caption{ \label{v3sfig2} 
{\it Left:} Trailed spectra of H$_{\beta}$: observed (left panel) and restored (middle panel) 
from the Doppler map for the H$_{\beta}$ line (right panel). {\it Right:} Doppler map for the H$_{\beta}$ line. Velocities at the axes are in 
1000 km s$^{-1}$ in the all panels.} 
\end{figure}

Doppler maps of the system  for six spectral emission lines  
were created using 
Spruit's code \citep{sp:98}. The trailed spectrogram for H$_{\beta}$ and corresponding Doppler map  
are shown in Fig.~\ref{v3sfig2}. It is clear from the Doppler map that there  
are two bright spots. The brighter spot corresponds to the meeting point of the  accretion stream with the disc and  
the second is in the opposite place of the disc.  
 
The obtained Doppler maps are very similar to those for IP Peg during quiescence  
\citep{ben:02}, and the two bright spots were interpreted as a two-arm spiral wave, where one spot is  
brighter due to interaction of the disc with the stream.  
3D hydrodynamical simulations \citep{bis:01}  predicted this  
picture and quasiperiodic brightness oscillations with period 
0.15 - 0.2 $P_{\rm orb}$, associated with the second arm rotation. We found the quasiperiodic brightness oscillations with a period $\approx 0.^d$01331 ($\approx  
0.19~ P_{\rm orb}$).  
 
\section {Estimation of system parameters} 
 
The mass of the secondary star can be evaluated from the orbital period using the 
mass-radius relation for main sequence stars: $M_{\rm RD} = 0.14 \pm 0.02 M_{\odot}$ \citep{knig:06, rap:01}.  
 
The eclipse width ($\Delta \varphi \approx$   0.025) provides us the relation between the inclination  
 $i$ and a mass ratio $q = M_{\rm RD}/ M_{\rm WD}$ \citep{horn:85}.  
 
We assume that the 
half of the distance between spectral line peaks corresponds to the Kepler velocity at the outer disc 
radius and obtain an additional relation between $M_{\rm RD}$, $i$ and $q$. 
The outer disc radius is limited by the tidal interaction with the 
secondary star \citep{pach:77} and we assume it to be equal to 0.8 radii of an 
equivalent Roche lobe volume sphere $R_{\rm L,WD}$ \citep{egg:83}. 
 
\begin{figure} 
  \includegraphics[height=.23\textheight]{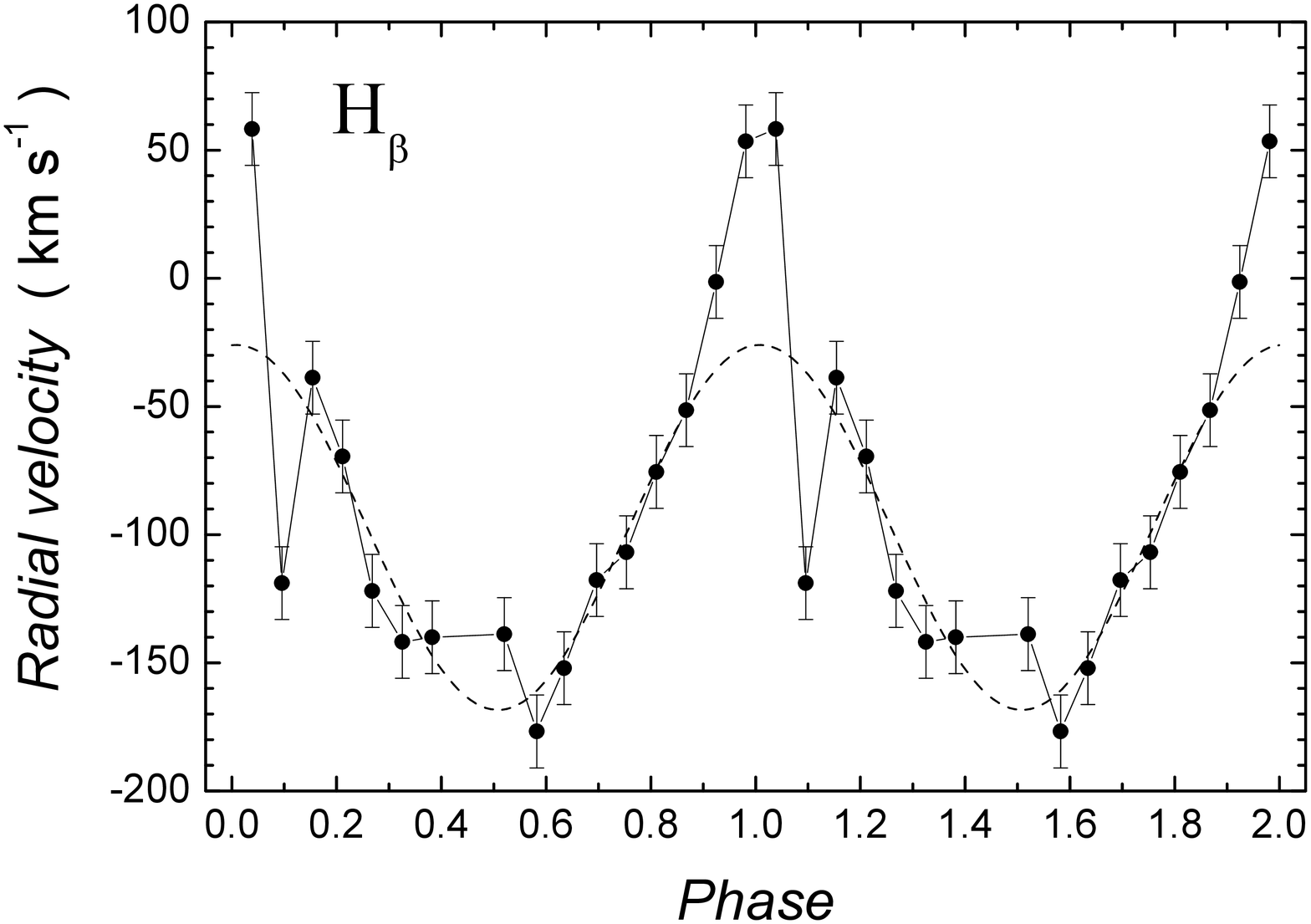} 
  \includegraphics[height=.23\textheight]{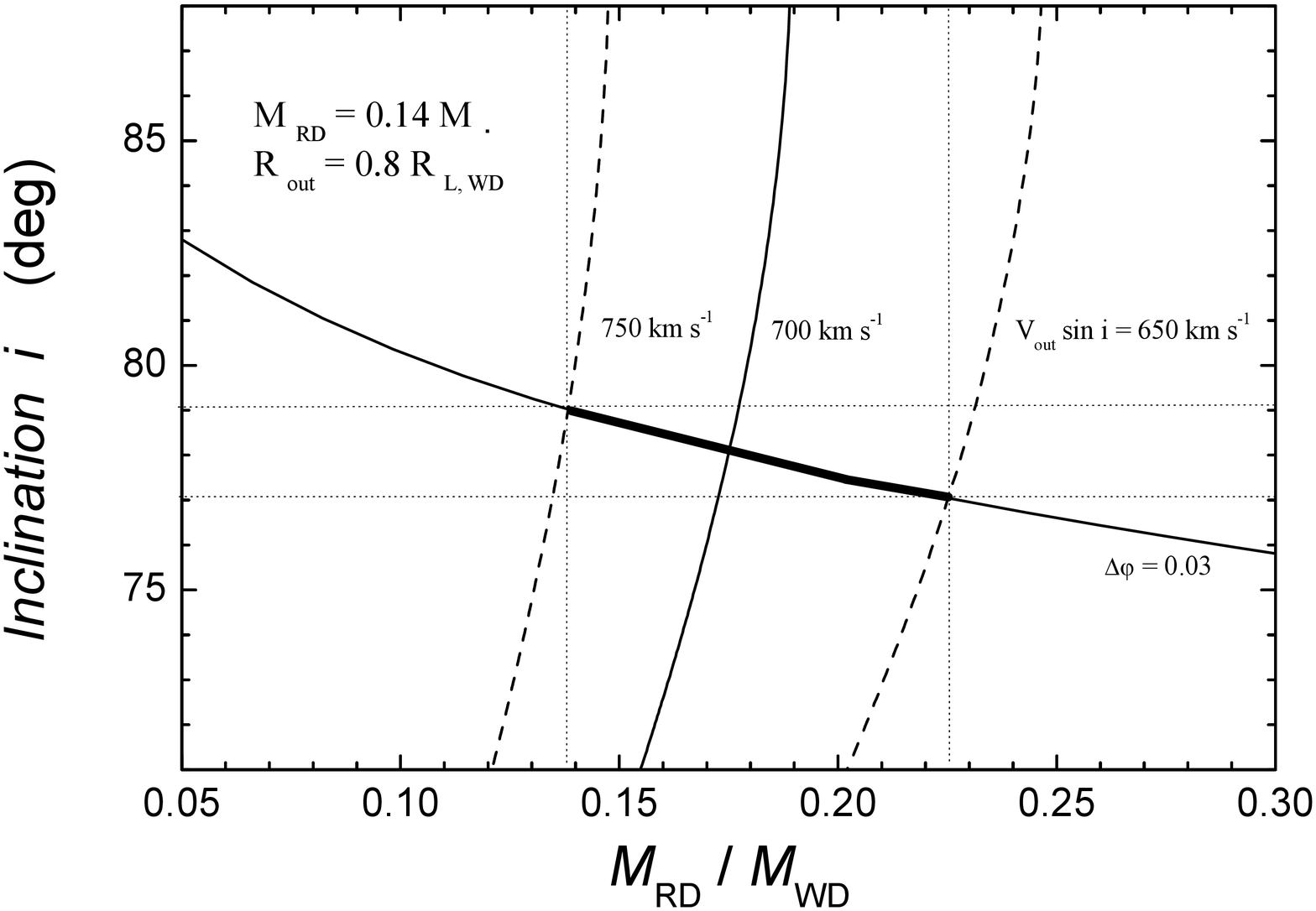} 
  \caption{ \label{v3sfig3} 
{\it Left:} Dependence of the radial velocity on the orbital phase, obtained using the 
H$_{\beta}$ emission line (filled circles). The sine, which fits the 
radial velocity curve in the 0.2 -- 0.9 phase range is shown by the dashed curve. {\it Right:} The relationship between the mass ratio $q$ and the inclination $i$ is 
shown for the phase width of the eclipce $\Delta~\varphi$ = 0.03 and for the three 
values of the observed velocity at the outer disc radius. The later 
relationships are calculated for $M_{\rm RD} = 0.14 M_{\odot}$. } 
\end{figure}

We measured the distance between peaks for the H$_{\beta}$ line at phase $\varphi \sim$ 0.5 and obtain $V_{\rm out} \sin i$ = 
700 $\pm$ 50 km s$^{-1}$.  Using this value we 
calculated the allowed region in the $i$--$q$ plane, shown in 
Fig.~\ref{v3sfig3}\,(right panel). The interception of the allowed region with the 
dependence $i$--$q$, found from the phase width of the eclipce, gives the allowed 
ranges of $i$ and $q$.  Finally, we have $M_{\rm WD} = 
0.8 \pm 0.22 M_{\odot}$ and $i = 78^{\circ} \pm 1.^{\circ}5$ for $M_{\rm 
RD} = 0.14 \pm 0.02~ M_{\odot}$, $V_{\rm out} \sin i$ = 700 $\pm$ 50 km 
s$^{-1}$ and $R_{\rm out} = 0.80 \pm 0.05~ R_{\rm L,WD}$.  
 
\section{Conclusions} 
 
We conclude that 1RXS\,J180834.7+101041 is a disc cataclysmic variable 
star with component masses  $M_{\rm WD} = 0.8 \pm 0.22 M_{\odot}$ and 
$M_{\rm RD} = 0.14 \pm 0.02~ M_{\odot}$, the inclination angle to line of 
sight $i = 78^{\circ} \pm 1.^{\circ}5$ and with a spiral density wave in 
the disc. Most probably it is an SU UMa type star, but more 
extended photometric and spectroscopic investigations have to be performed 
to confirm this suggestion. 
 
\begin{theacknowledgments} 
   This work is supported by the 
Russian Foundation for Basic Research (grant  09-02-97013-p-povolzhe-a), and the DFG SFB\,/\,Transregio 7 ``Gravitational Wave Astronomy'' (V.S.). 
\end{theacknowledgments} 

\bibliography{v3suleimanov}
\bibliographystyle{aipproc}

\end{document}